%% file: twozonemod.tex
\documentclass[fleqn]{annalenmod}

\usepackage{graphics}
\usepackage{graphicx}
\usepackage{amsmath}
\usepackage{amssymb}

\begin{document}
%%%%%%%%%%%%%%%%%%%%%%%%%%%%%%%%%%%%%%%%%%%%%%%%%%%%%%%%%%%%%%%%%%%%%%%%%%%%%%
%%%%%%%% the following newcommands will be completed by the publisher %%%%%%%%
%%%%%%%%%%%%%%%%%%%%%%%%%%%%%%%%%%%%%%%%%%%%%%%%%%%%%%%%%%%%%%%%%%%%%%%%%%%%%%
\newcommand{\volume}{11}              %sets current volume,
\newcommand{\xyear}{2000}            %sets year in header
\newcommand{\issue}{5}               %sets current issue,
\newcommand{\recdate}{15 November 1999}  %sets received date,
\newcommand{\revdate}{dd.mm.yyyy}    %sets revised date,     
\newcommand{\revnum}{0}              %number of revisions,
\newcommand{\accdate}{dd.mm.yyyy}    %sets accepted date,
\newcommand{\coeditor}{ue}           %sets (co)editor,
\newcommand{\firstpage}{1}         %first page number,  
\newcommand{\lastpage}{4}          %last page number,
\setcounter{page}{\firstpage}        %sets page counter to first page number 

\newcommand{\keywords}{general relativity, relativistic stars,
  gravitational waves} 

\newcommand{\PACS}{04.30.Db, 04.40.Dg}

\newcommand{\shorttitle}
{Karlovini et al., Compact stellar objects with multiple neck optical geometries} 

%%%%%%%%%%%%%%%%%%%%%%%%%%%%%%%%%%%%%%%%%%%%%%%%%%%%%%%%%%%%%%%%%%%%%%%%%%%%%%
%                           Kjell's stuff                                    %
%%%%%%%%%%%%%%%%%%%%%%%%%%%%%%%%%%%%%%%%%%%%%%%%%%%%%%%%%%%%%%%%%%%%%%%%%%%%%%

\newcommand{\acronym}[3]{\newcommand{#1}{#3 (#2)\relax\renewcommand{#1}{#2}}}
\newcommand{\NS}{N_{\rm Schw}} % Schwarzschild radial gauge function
\newcommand{\RTaub}{{\hat r}}
\newcommand{\Radial}{{\tilde r}}
\newcommand{\ME}{\Mscr_{\rm E}} % The embedding cylinder
\newcommand{\BE}{\begin{eqnarray}}
\newcommand{\EE}{\end{eqnarray}}

% Blackboard bold characters
\newcommand{\Dbb}{{\Bbb D}}  \newcommand{\Rbb}{{\Bbb R}}

% Acronyms
\acronym{\SSS}{SSS}{{\em static spherically symmetric}}
\acronym{\TOV}{TOV}{{\em Tolman-Oppenheimer-Volkoff}}
\acronym{\NMU}{NMU}{{\em Nilsson-Marklund-Uggla}}

\include{journabr}
\newcommand{\eprint}{\textsf}

\title{Compact stellar objects with multiple neck optical geometries}

\author{M.\ Karlovini$^{1}$, K.\ Rosquist$^{1}$, and L.\ Samuelsson$^{1}$} 

\newcommand{\address}
  {$^{1}$Department of Physics, Stockholm University 
  \\ \hspace*{0.5mm} Box 6730, 113 85 Stockholm, Sweden} 

\newcommand{\email}{\tt kr@physto.se} 
\maketitle

\begin{abstract}
Ultracompact stellar models with a two-zone uniform density equation
of state are considered. They are shown to provide neat examples of
optical geometries exhibiting double necks, implying that the
gravitational wave potential has a double well structure.
\end{abstract}

\section{Introduction}
The \SSS\ relativistic stellar models have been the subject of extensive
studies in the past (see for example \cite{dl:sss} for a review of exact
perfect fluid \SSS\ solutions of the Einstein equations up to 1994).  De
Felice \cite{defelice:trapped} first discussed the possibility of trapping of
massless particles in the interior of compact stellar objects.  Abramowicz and
coworkers \cite{aabgs:optical} considered the optical geometry of uniform
density models.  They found that the optical geometry (for geodesics in the
equatorial plane) of the stellar interior was identical to a standard
2-sphere.  They also showed that the optical geometry of the uniform density
models with $R<3M$ had a neck (at $R=3M$) and a corresponding bulge in the
interior.  The bulge is associated with stable spatially closed orbits of
massless (or ultrarelativistic) particles and waves propagating at the speed
of light (in the geometric optics approximation).  This provided a nice visual
illustration of the phenomenon of trapping of waves and matter moving at
relativistic velocities inside compact objects.  Moreover, it allowed a simple
derivation of the eigenfrequencies of the w-modes of trapped gravitational
waves.

Despite the relative simplicity of the field equations for static spherically
symmetric perfect fluid models, the full picture of the structure of the
gravitational field of such objects is only now beginning to emerge.
Following the powerful Nilsson-Uggla formulation of the \SSS\ field
equations \cite{nmu:sss_phase}, a novel feature which was recently uncovered
is the existence of optical geometries with multiple necks
\cite{rosquist:multneck}.  One might naively expect that such behavior would
only occur for rather strange or unphysical equations of state.  However, as
shown in \cite{rosquist:multneck}, multiple necks do in fact appear for the
innocent looking Zel'dovich stiff matter equation of state with an added
constant $p=\rho-\rho_{\rm s}$.  We are using a subscript ``s'' to denote
evaluation at zero pressure ({\em i.e.\/} at the stellar surface).  In fact if
$n$ is the number of the necks, then $n \rightarrow \infty$ for stellar models
with that equation of state in the limit of infinite central density, $\rho_{\rm
c} \rightarrow \infty$, where the subscript ``c'' is used for evaluation at
the center of the stellar object.  Actually, most of these models correspond
to unstable equilibria and so would not be realizable as stable stellar
objects.  However, unstable equlibria can play a role as intermediate states
in gravitational collapse situations \cite{nc:criticalperfect}.

The Zel'dovich equation of state has the advantage of being causal in the
sense that the adiabatic speed of sound equals the speed of light, $v_{\rm
sound} := \sqrt{dp/d\rho} =1$.  However, from the point of view of visualizing
the multiple neck optical geometry, it is not so well suited because of the
rather small amplitude of the necks for that equation of state.  Therefore, to
obtain better visual displays we have used double layer uniform density models
to exhibit double neck optical geometries. Such models were used by
Lindblom \cite{lindblom:phase} as prototypes to discuss phase
transitions in relativistic stellar objects.

\section{Some multiple neck optical geometries}
\label{sec:multneck}

The metric for a \SSS\ spacetime can be
written in the Schwarzschild radial gauge as
\BE
ds^2=-e^{2\nu}dt^2+e^{2\lambda}dr^2+r^2(d\theta^2+\sin^2\theta d\phi^2),
\EE
where $\nu$ and $\lambda$ are functions of $r$. 
When studying null geodesics, it is often convenient to
introduce the {\em optical geometry} of the spacetime, defined by
\BE
d\tilde{s}^2 = e^{-2\nu}ds^2 = -dt^2 + e^{2(\lambda-\nu)}dr^2 +
e^{-2\nu}r^2d\phi^2.
\EE
Since this metric is conformally related to the physical metric, the
spacetime and its optical geometry have the same null geodesics. 
Due to the symmetries of the spacetime we may, without loss of generality, 
consider only the spacelike slice given by $t=0$ and
$\theta=\pi/2$. Introducing the tortoise radial variable defined by
$dr_*=e^{\lambda-\nu}dr$, the metric on this slice may be written
\BE
dl^2=dr_*^2+\tilde{r}^2d\phi^2,
\EE
where $\tilde{r}=e^{-\nu}r$.
In order to gain intuitive understanding of this geometry it is
useful to embed it in 3-dimensional Euclidean space. Standard embedding 
techniques ({\em c.f. e.g.} \cite{mtw:gravitation}) yield the differential equation 
for the embedded surface
\BE
\tilde{r}=e^{-\nu}r, & &
h'=e^{-\nu}\sqrt{e^{2\lambda}-(1-r\nu')^2},
\EE
where $\tilde{r}$ and $h$ are the radius and height coordinates in a
standard cylindrical coordinate system and a prime denotes
differentiation with respect to $r$. 

As shown by Chandrasekhar and Ferrari \cite{cf:osc} the axial modes of nonradial 
metric perturbations of SSS perfect fluids are (to first order) not
coupled to a perturbation of the matter. Such metric perturbations may
therefore be interpreted as pure gravitational waves. By separation of
variables the equations governing these modes can be reduced
to the 1-dimensional Schr\"odinger equation  
\BE
  -\frac{d^2}{dr_{\!*}^2}Z_{l} + (V_{l}+V)Z_{l} = \sigma^2 Z_{l}
\EE
where the total potential $V_{l}+V$ is dominated by $V_{l} =
l(l+1)\tilde{r}^{-2}$, $l=2,3,\ldots$ which is simply the effective potential for null
geodesics with angular momentum $L^2 = l(l+1)$. 
This shows that the optical geometry is closely related to the trapped
modes of gravitational radiation since the maxima and minima of
$V_l$ clearly correspond to the necks and bulges of the optical
geometry respectively. 
We turn now to the analysis of the optical geometry of the double layer
uniform density models. The equation of state is parametrized by the
constant densities $\rho_{+}$ and $\rho_{-}$ of the interior and
exterior zone respectively, as well as the transition pressure
$p_{\mathrm t}$ where the density is discontinous. Given these
parameters, a stellar model is obtained by specifying the central
pressure $p_{\mathrm c}$. Below we display the optical geometry and
the axial mode potentials for three stars of this type. The parameter
values are chosen to obtain three distinct examples of optical
geometries with pronounced double necks, corresponding to two deep
potential wells. 
As the wells are fairly rectangular-shaped with a
certain width $L$ and height $V_0$, one can make a crude estimate of 
the number of trapped modes by taking $Z_l$ to be the wave
function of an infinite square well, with the same width $L$, and
count the number of states with energy less than $V_0$. The number of
trapped modes is then approximately given by $\sqrt{V_0}L/\pi$.
  
The models considered here were chosen to have the same value of the tenuity
$\alpha=R/M=2.2749$, and the overall scale was fixed by setting the
total mass equal to unity (in units such that $G_{ab}=T_{ab}$). Below the optical
geometries are displayed next to the potentials relevant for the axial 
modes with $l=2$, the solid line indicating the total potential ($V+V_l$), the
dashed line showing
$V_l$ and the dash-dotted line showing $V$. In the optical geometry the two
shades of gray indicate the different density layers, and the white part
corresponds to the exterior Schwarzschild geometry. The parameter
values, given in units of the appropriate power of the total mass, are
given in the respective captions. We may remark however that none of
these models are likely to be stable, but that we have found models with less
pronounced double necks that are probably stable.

\begin{figure}[ht]
 \begin{minipage}[b]{0.4\linewidth}
  \centering
  \includegraphics[scale=.55]{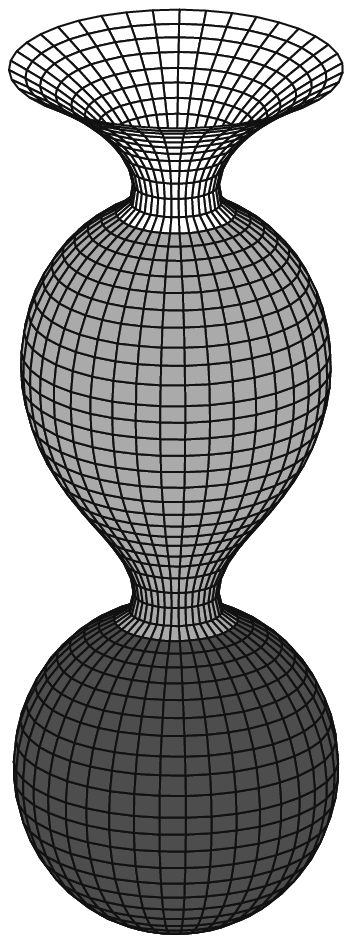}
 \end{minipage}%
 \begin{minipage}[b]{0.6\linewidth}
  \includegraphics[scale=.37]{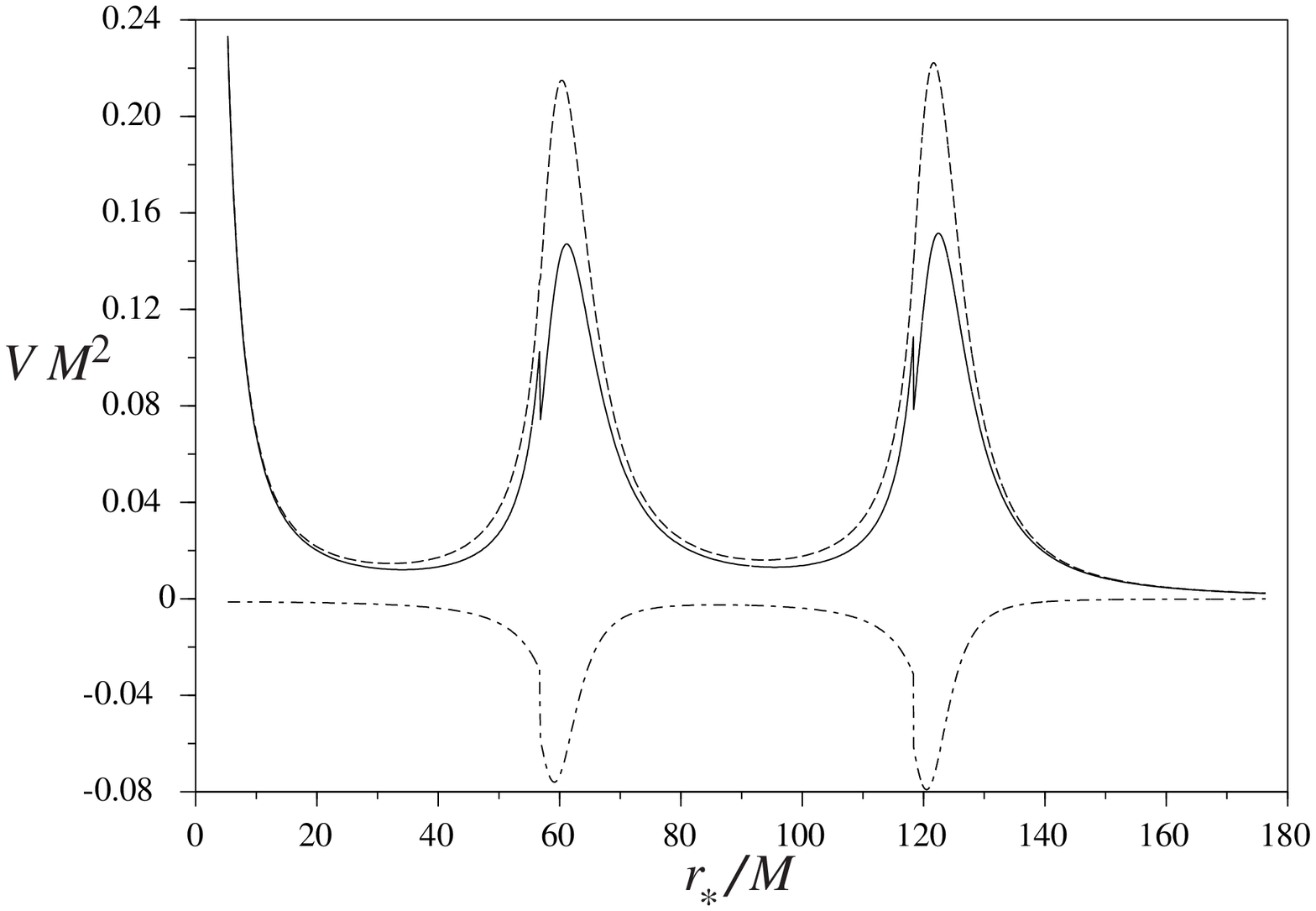}
 \end{minipage}
 \caption{The ``peanut'' geometry and the corresponding
   potentials. The parameter values are $\rho_+=2.512\times 10^3$,
   $\rho_-=0.5024$, $p_t=35.89$, and $p_c=5.778\times 10^4$.}
 \label{fig:peanut}
\end{figure}
\begin{figure}[ht]
 \begin{minipage}[b]{0.4\linewidth}
  \centering
  \includegraphics[scale=.55]{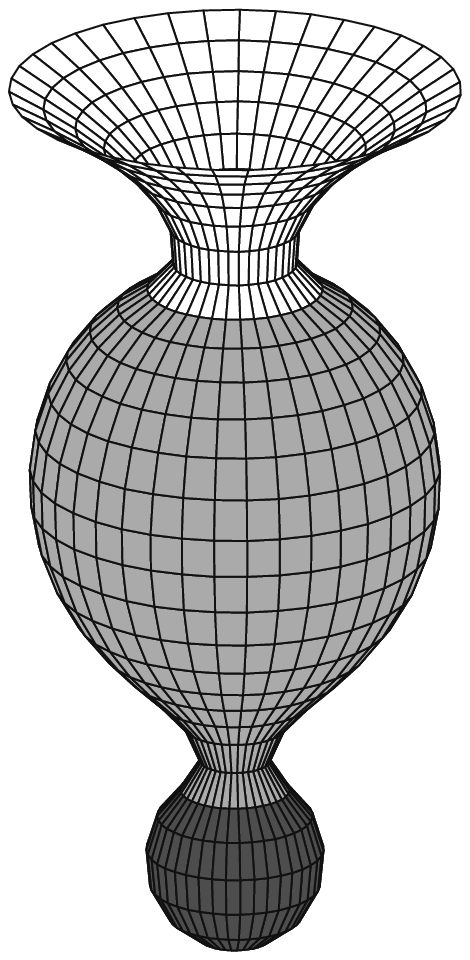}
 \end{minipage}%
 \begin{minipage}[b]{0.6\linewidth}
  \includegraphics[scale=.37]{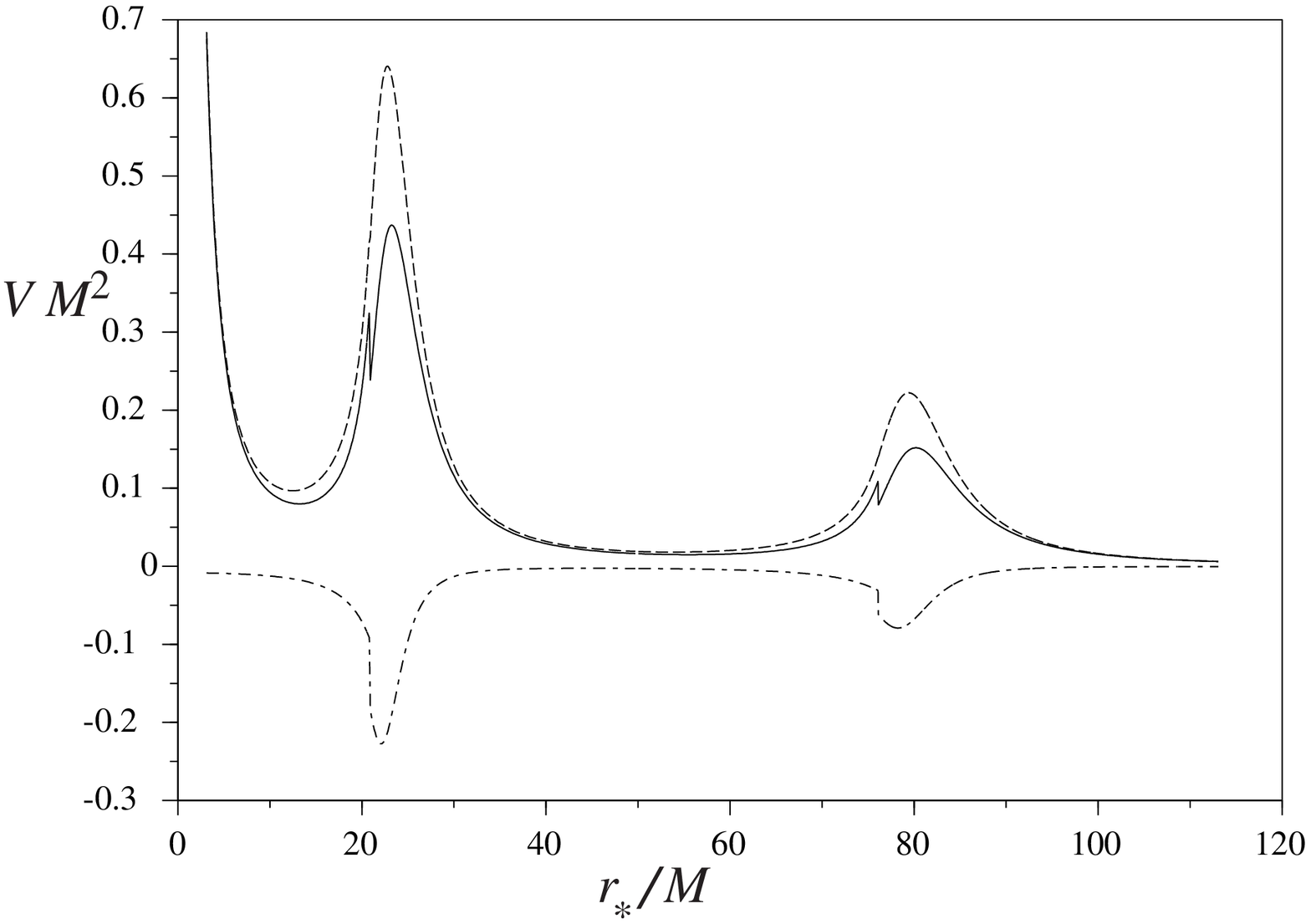}
 \end{minipage}
 \caption{The ``bigup'' geometry and the corresponding potential. The
   parameter values are $\rho_+=5.325\times 10^3$,
   $\rho_-=0.5047$, $p_t=29.58$, and $p_c=5.325\times 10^4$.}
 \label{fig:bigup}
\end{figure}
\begin{figure}[ht]
 \begin{minipage}[b]{0.4\linewidth}
  \centering
  \includegraphics[scale=.55]{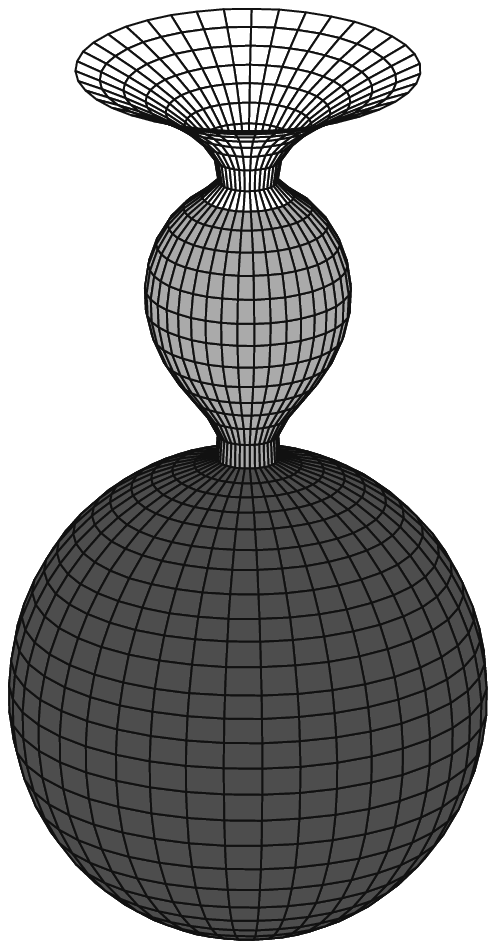}
 \end{minipage}%
 \begin{minipage}[b]{0.6\linewidth}
  \includegraphics[scale=.37]{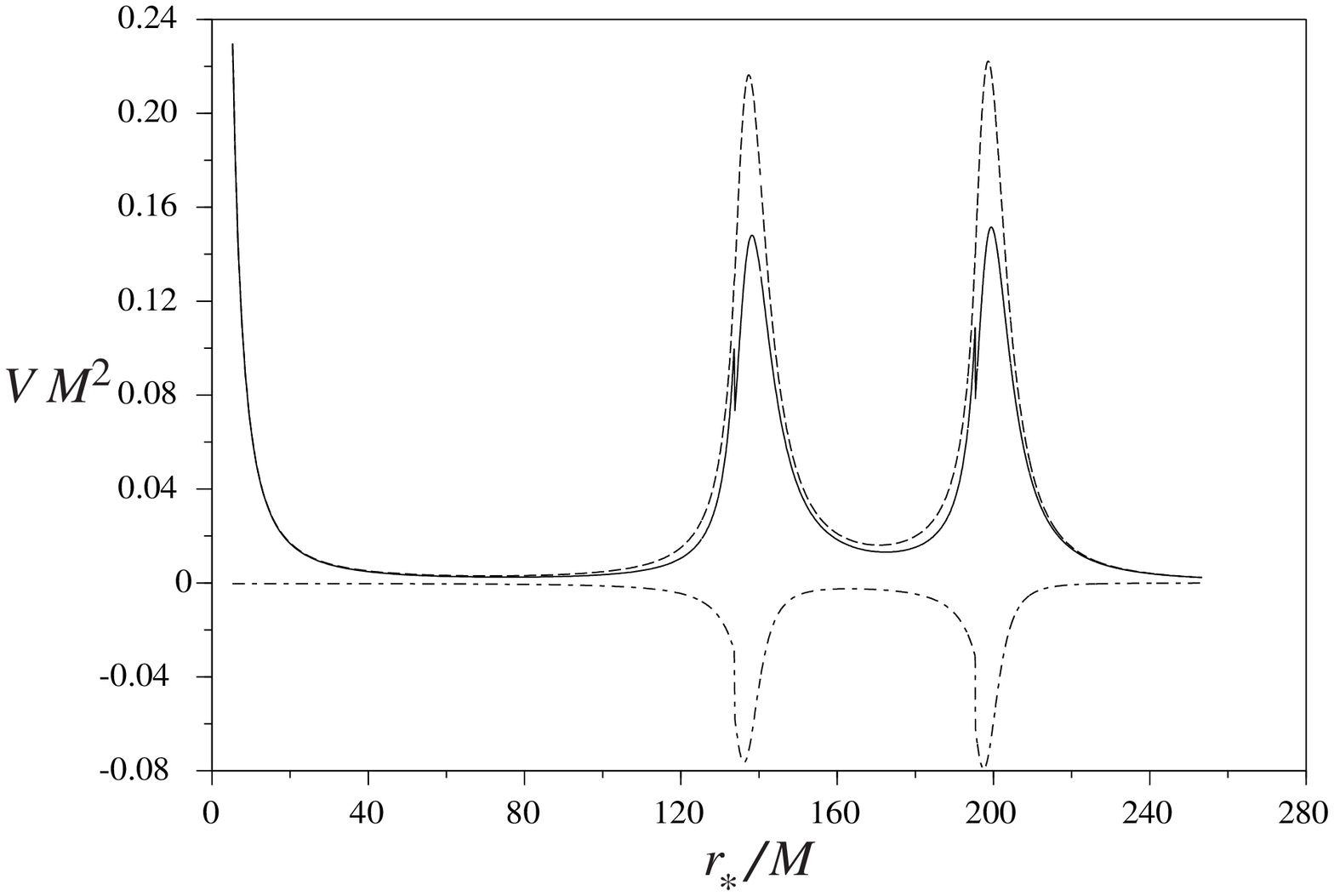}
 \end{minipage}
 \caption{The ``bigdown'' geometry and the corresponding potential. The
   parameter values are $\rho_+=2.569\times 10^3$,
   $\rho_-=0.5024$, $p_t=36.69$, and $p_c=2.954\times 10^5$.}
 \label{fig:bigdown}
\end{figure}

\end{document}

%% file: journabr.tex
% some journal abbreviations (mostly physics related):
\def\journalfont{\rm}         % this allows redefinition of the font later
\def\jou#1{{\journalfont #1\ }}
\def\joudef#1#2{\def #1{\jou{\ignorespaces #2}}}

\joudef{\aaa}    { Astron.\ Astrophys.}
\joudef{\aip}    { Adv.\ Phys.}
\joudef{\adm}    { Adv.\ Math.}
\joudef{\am}     { Ann.\ Math.}
\joudef{\apny}   { Ann.\ Phys.\ (N.Y.)}
\joudef{\apj}    { Astrophys.\ J.}
\joudef{\cjp}    { Can.\ J.\ Phys.}
\joudef{\cmp}    { Commun.\ Math.\ Phys.}
\joudef{\cqg}    { Class.\ Quantum Grav.}
\joudef{\grg}    { Gen.\ Rel.\ Grav.}
\joudef{\ijmpd}  { Int.\ J.\ Mod.\ Phys.\ D}
\joudef{\ijtp}   { Int.\ J.\ Theor.\ Phys.}
\joudef{\invm}   { Invent.\ Math.}
\joudef{\jm}     { J.\ Math.}
\joudef{\jmaa}   { J.\ Math.\ Anal.\ Appl.}
\joudef{\jmp}    { J.\ Math.\ Phys.}
\joudef{\jpa}    { J.\ Phys.\ A}
\joudef{\mnras}  { Mon.\ Not.\ R.\ Ast.\ Soc.}
\joudef{\mpla}   { Mod.\ Phys.\ Lett.\ A} 
\joudef{\nature} { Nature}
\joudef{\nc}     { Nuovo Cim.}
\joudef{\npb}    { Nuc.\ Phys.\ B}
\joudef{\ph}     { Physica}
\joudef{\pla}    { Phys.\ Lett. A}
\joudef{\plb}    { Phys.\ Lett. B}
\joudef{\pr}     { Phys.\ Rev.}
\joudef{\prd}    { Phys.\ Rev.\ D}
\joudef{\prep}   { Phys.\ Rep.}
\joudef{\prl}    { Phys.\ Rev.\ Lett.}
\joudef{\prsla}  { Proc.\ Roy.\ Soc.\ Lond.\ A}
\joudef{\ptp}    { Prog.\ Theor.\ Phys.}
\joudef{\ptps}   { Prog.\ Theor.\ Phys.\ Suppl.}
\joudef\rmp      { Rev.\ Mod.\ Phys.}
\joudef\spj      { Sov.\ Phys.\ JETP}
%%%%%%%%%%%%%%%%%%%%%%%%%%%%%%%%%%%%%%%%%%%%%%%%%%%%%%%%%%%%%%%%%%%%%%%%%%%%%%